\documentclass[conference]{IEEEtran}
\IEEEoverridecommandlockouts

\usepackage{comment}

\usepackage{booktabs}
\usepackage{multirow}

\usepackage{textcomp}

\usepackage{cite}

\ifCLASSINFOpdf
\usepackage{graphicx}
\usepackage{siunitx}
 \usepackage{booktabs}
 \usepackage{orcidlink}

\else
\fi

\begin{document}
%
\title{On the Design of a Slow Control Interlock System for the European GateMate FPGA}




%
\author{
\IEEEauthorblockN{
Vitalii Burtsev\protect\orcidlink{0000-0003-0858-5384},
Daniele Passaretti\protect\orcidlink{0000-0001-7154-8354},
Daniel Anders\protect\orcidlink{0009-0003-1853-1133},
Christof Pfannenmüller\,\protect\orcidlink{0009-0006-2370-626X},
Fabian Lurz\,\protect\orcidlink{0000-0003-4948-9655},
Thilo Pionteck\,\protect\orcidlink{0000-0001-6518-1226}
}
\IEEEauthorblockA{
Otto-von-Guericke University Magdeburg, Institute for Information Technology and Communications, Germany\\
Email: name.surname@ovgu.de
}
}

\maketitle

\begingroup
\renewcommand{\thefootnote}{}
\footnotetext{%
\textit{This paper has been accepted at the
2026 IEEE Nordic Circuits and Systems Conference (NorCAS 2026).}

\par\smallskip
\noindent
\copyright\ 2026 IEEE. Personal use of this material is permitted.
Permission from IEEE must be obtained for all other uses,
in any current or future media, including reprinting/republishing
this material for advertising or promotional purposes,
creating new collective works, for resale or redistribution
to servers or lists, or reuse of any copyrighted component
of this work in other works.
}
\endgroup

\begin{abstract}
Slow Control Systems (SCSs) are essential for the monitoring, configuration, and protection of distributed safety-critical systems in scientific, industrial, and medical applications. Although software-based SCS architectures offer flexibility and scalability, they suffer from software aging and non-deterministic latencies, which limit long-term reliability and timely reaction to critical errors, often requiring additional hardware interlock systems.
This paper presents an FPGA-centric lightweight Slow Control Interlock System (SCIS) designed to achieve deterministic ultra-low-latency reaction times in long-running dependable systems. System parameters are acquired and fused at runtime in a Supervisory Monitor Unit (SMU), enabling real-time fault condition classification and prioritization through configurable priority rules and triggering kill-switch actions with microsecond-scale latency.
The proposed SMU has been described in VHDL as vendor-agnostic architecture, and assessed on the 28-nm GateMate M1A1 FPGA, using the open-source OSS-CAD-Suite toolchain.
Experimental results show that the SMU architecture utilizes only 37,6\% of LUTs, 9,4\% of FFs, and 35,9\% of BRAM.
Furthermore, this low resource utilization allow to achieve 40,48 ms reconfiguration time, using OSS-CAD-Suite including gmpack v1.13. This is a key metric for rejuvenation and consequently for enabling aging mitigation in critical applications. Offloading the Ethernet stack to the W5500 Ethernet module and integrating the Grafana dashboard provides a lightweight, deterministic SCIS. This demonstrates the feasibility of our FPGA-centric architecture as a solution to the problem of combining ultra-low-latency interlock capabilities with the scalability of software-based systems.

\end{abstract}

\IEEEpeerreviewmaketitle

\section{Introduction}

Field-Programmable Gate Arrays (FPGAs) are widely adopted in safety- and time-critical subsystems due to their deterministic execution, ultra-low latency,  and intrinsic isolation support~\cite{alachiotis2025fpga,10.1007/978-3-031-21867-5_10}. These properties make them particularly suitable for real-time data collection and processing, machine protection and interlock operations in high-energy physics (HEP) experiments and medical systems, such as experiments on particle accelerators and Magnetic Resonance Imaging (MRI) platforms~\cite{guo2024studies,keyes2017development,stillman1995slow,LI2025103345,4448450,7097454,6829202,zappa1607versatile,8977848}.

In such systems, FPGA-based designs are commonly used for detector front-end electronics, data acquisition, and interlock subsystems, while configuration, monitoring, and management are handled by software-based Slow Control Systems (SCSs)~\cite{bai2017intelligent}. Although this separation simplifies development, it introduces key limitations. Software-based SCSs suffer from aging-related reliability degradation and often require reboots, leading to unavoidable downtime~\cite{6698905}. This is especially problematic in safety-critical medical environments such as MRI systems, where continuous monitoring of system parameters is essential and shutdowns are costly, time-consuming, and may pose safety risks due to helium loss and magnet quenching~\cite{lisanti2018mri, MHRA2021}. Furthermore, separating interlock logic from monitoring infrastructure limits the availability of time-correlated diagnostic data for interlock systems, hindering effective fault analysis and predictive maintenance~\cite{Huovinen2010LargeScale,Waszecki2016SystemLevel}.

Recent low-cost FPGA platforms offer an opportunity to revisit this traditional separation. Reliable and simplified microarchitectures, competitive logic density, and increasing support for open-source toolchains enable the implementation of lightweight reconfigurable architectures that integrate both protection and monitoring functionalities within a deterministic hardware environment. 
However, it remains unclear to what extent FPGA-based designs are suitable for tasks that do not require strict time constraints, such as SC systems, and what trade-offs they incur, especially during long-term operation in safety-critical environments. Typical examples include experiments in high-energy physics, such as ATLAS~\cite{keyes2017development}, AGS E-821~\cite{stillman1995slow}, the cooling storage ring~\cite{LI2025103345}, OPERA~\cite{4448450}, and Belle II~\cite{7097454}, as well as smaller-scale medical systems, such as PET/MRI platforms~\cite{6829202,zappa1607versatile}.

This work proposes an FPGA-centric Slow Control Interlock System (SCIS) that integrates monitoring and machine protection logic within a lightweight FPGA-based architecture. The system collects and monitors measurements of system parameters such as voltages or temperatures, coming from different subsystem components having different custom and standard communication protocols, classifies devices' priorities according to predefined severity policies, and triggers interlock actions with ultra-low latency. Diagnostic information is exported through a minimal hardware–software interface and visualized using standard monitoring frameworks such as Grafana dashboard~\cite{GrafanaDashboards}.
A key contribution of this paper is the evaluation of emerging low-cost FPGA platforms and its open-source design flow for implementing the proposed architecture. 
In particular, the GateMate FPGA was selected because it combines a low cost per lookup table (LUT) with radiation hardness comparable to that of mature FPGA families, such as the AMD-Xilinx 7 series~\cite{koers2024assessing}. In addition to these technical advantages, this choice aligns with Europe’s renewed emphasis on technological sovereignty, reaffirmed by the EU’s “Technological Sovereignty Package” and the proposed “Microchips Act” 2.0 of June 2026 that aim to open-source design flow within EU supply chain~\cite{ec2026sovereignty}.

The GateMate is developed and manufactured in Germany by Cologne Chip~\cite{colognechip_gatemate1_datasheet} and by Globalfoundries respectively, and its fully open-source flow keeps the entire RTL-to-bitstream path auditable and reproducible, free of vendor IP and license requirements. For protection systems embedded in medical and scientific infrastructure with decade-scale lifetimes, this yields concrete benefits: inspectable generation of the safety logic, independence from export-control and licensing discontinuities, and the ability to rebuild bitstreams throughout the system's service life.

The main contributions of this work are:
\begin{itemize}
    \item A lightweight FPGA-centric architecture integrating slow-control monitoring and interlock protection in a single fabric, suitable for MRI, open-source available on GitHub\cite{HTI-SCIS}.
    \item An experimental evaluation of the GateMate FPGA as a low-cost reconfigurable platform for control and error monitoring.
    \item A practical assessment of the open-source OSS-CAD-Suite toolchain for the GateMate FPGA.
\end{itemize}

The remainder of this paper is organized as follows. Section~II introduces interlocks and SCSs, and the GateMate FPGA. Section~III presents the proposed SCIS. Section~IV reports the GateMate FPGA experience. Section~V discusses the experimental results. Section~VI concludes the paper.

\section{Background}

\subsection{Slow Control Systems and Interlock System}
Slow Control Systems (SCS) monitor and regulate operational parameters such as power supply voltage, room temperature, gas flows in medical devices and nuclear physics experiments, operating on timescales much slower than detector data streams~\cite{keyes2017development,stillman1995slow,LI2025103345,4448450,7097454,6829202,zappa1607versatile}. These systems must integrate system parameters from numerous different subsystems, often supplied by different hardware vendors, requiring scalable and vendor-agnostic architectures.

In common designs, monitoring and control functions are separated from safety-critical Interlock mechanisms to avoid interference with fast kill-switch actions. While this facilitates safety, it limits real-time correlation between detected faults and contextual monitoring data. Existing solutions are largely software-based and distributed~\cite{keyes2017development,stillman1995slow,LI2025103345,4448450,7097454,6829202,zappa1607versatile}, limited also by the rejuvenation time, which is usually in the order of minutes~\cite{6698905}. In contrast, this work proposes a new SCS and Interlock design that improves integration between monitoring data and safety logic without compromising real-time protection. 

\subsection{GateMate FPGA}

\begin{figure}[!ht]
    \centering
    \includegraphics[width=0.9\linewidth]{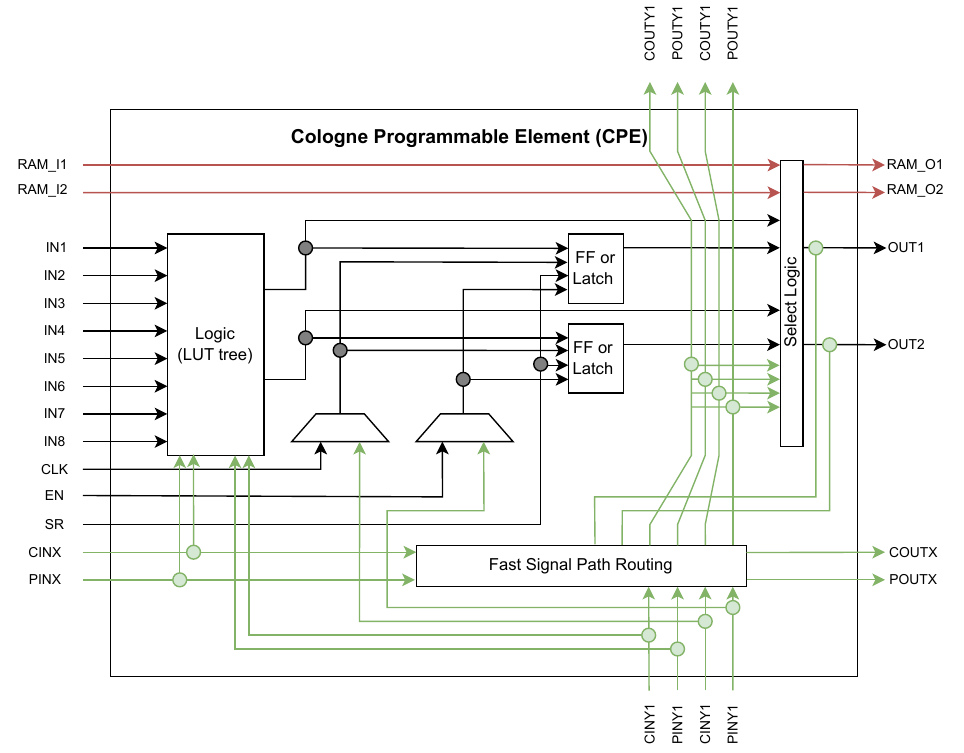}
    \caption{Cologne Programmable Element~\cite{colognechip_gatemate1_datasheet}.}
    \label{fig:CPEfigure}
\end{figure}

The  GateMate FPGA family is completely designed and manufactured in Europe.
The core of this FPGA family is the CPE block shown in Fig.~\ref{fig:CPEfigure}: a configurable logic block with an 8-input LUT which can be treated by placer as two 4-input LUT trees, two memory elements (a flip-flop or latch), and fast carry chains, which enables the efficient implementation of combinational and arithmetic logic~\cite {colognechip_gatemate1_datasheet}.

In the GateMate CCGM1A1 FPGA, CPEs are arranged in a 160x128 matrix, including 32 dual-port Block RAM (BRAM) cells. 
Each physical 40K BRAM tile is separable into two independent 20K BRAM blocks, usable in Simple Dual Port or in True Dual Port  mode.
A Phase-Locked Loop (PLL) inside the CCGM1A1 FPGA provides up to four system clock signals for the implemented design.

The CCGM1A1 FPGA Chip can operate in the temperature range of \SI{-40}{\celsius} up to \SI{125}{\celsius} and Total Ionizing Dose (TID) tests at the CERN Highly-Accelerated Mixed Field Facility (CHARM) have shown that this FPGA is robust against degradation up to $\SI{1}{\kilo Gy}$ = $1000 \frac{\si{\joule}}{\si{\kg}}$ of ionizing radiation, which makes it not only suitable for safety critical applications like MRI with harsher environmental conditions, but also for nuclear applications such as PET physics and HEP experiments~\cite{koers2024assessing}.

\subsection{Design flow}
\begin{figure}[!ht]
    \centering
    \includegraphics[width=1\linewidth]{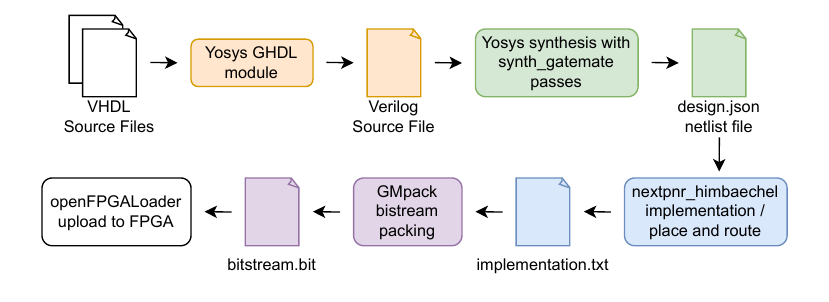}
    \vspace{-8mm}

    \caption{OSS-CAD-Suite workflow for the GateMate FPGA.}
    \label{fig:Workflow}
\end{figure}

\begin{figure*}[!tbp]
\centering
\includegraphics[width=0.8\textwidth]{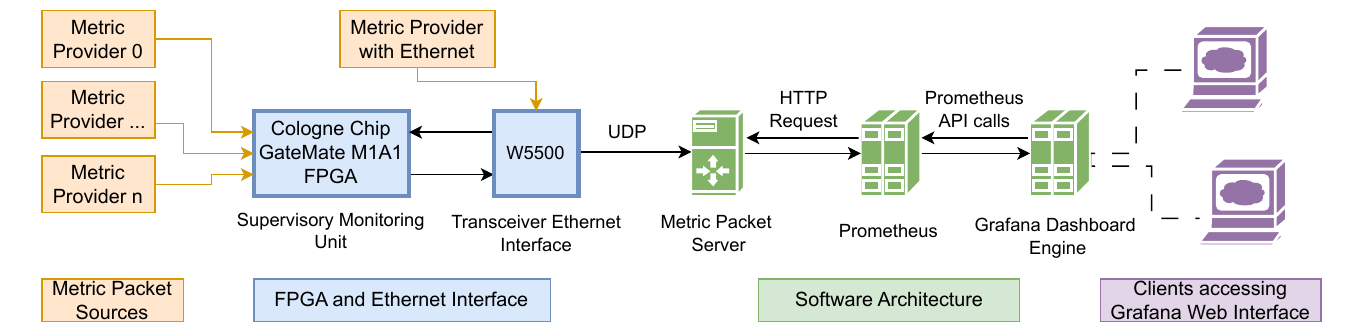}
\caption{Overview of SCIS architecture.}
\label{fig:InfrastructureOverview}
\end{figure*}

The adopted design flow follows the current reference workflow for the GateMate  FPGA family and combines open-source tools such as GHDL, Yosys, nextpnr and openFPGAloader as recommended by the device vendor.

The GateMate FPGA uses the OSS-CAD-Suite, an open-source software distribution used in digital logic design and maintained by YosysHQ~\cite{yosys_v0_53}. 
It provides the toolchain shown in Fig. \ref{fig:Workflow}, that enables rapid development iterations with full visibility into resource utilization and timing analysis.

Yosys requires the GHDL plugin to convert VHDL into a compatible format before synthesis. To target the CCGM1A1 FPGA, Yosys invokes the \texttt{synth\_gatemate} command, a script that executes various passes, including logic optimization, dead-code elimination, and technology mapping, to generate the final netlist.
Nextpnr, an open-source place and route tool, requires Yosys to emit a JSON netlist file.
Nextpnr-himbaechel is a modified version of nextpnr incorporating the work of project peppercorn, that allows the synthesized design to be mapped towards the CCGM1A1 FPGA.
As the implementation tool only outputs the implemented design as a human-readable file, it needs to be packed as a bitstream using GMpack before programming the Cologne Chip Evaluation Board using openFPGALoader.
The GateMate M1A1 Evaluation board features an integrated FT2232H USB interface chip, which provides dual channel USB-JTAG/SPI connectivity and enabling direct bitstream upload from a host computer via a single USB connection.
Bitstream uploaded using the SPI mode is written towards a Macronix MX25R6435F \SI{64}{Mb} flash memory, achieving millisecond-level FPGA reconfiguration time, depending on the size of the bitstream.
\section{Slow Control Interlock System Architecture}

The proposed FPGA-centric SCIS architecture shown in Fig.~\ref{fig:InfrastructureOverview} comprises seven functional entities: the \textbf{Metric Provider Source (MPS)}, where system parameters are generated; the \textbf{Supervisory Monitor Unit (SMU)}, a lightweight FPGA-based core that monitors system parameters and toggles kill-switch actions based on predefined rules; the \textbf{Transceiver Ethernet Interface (TEI)}, which offloads the Ethernet stack by bridging SPI-based metric packets to UDP/IPv4; the \textbf{Metric Packet Server (MPSrv)}, facilitating socket-based communication between the TEI and the database; \textbf{Prometheus}, a time-series database for real-time storage; the \textbf{Grafana Dashboard Engine (GDE)}, which processes stored data; the \textbf{Grafana Dashboard Visualizer (GDV)}, providing an end-user interface for parameters and alerts visualization.

The SCIS architecture, as shown in Fig.~\ref{fig:InfrastructureOverview}, scales by increasing the number of MPSs and supports multiple users through additional GDV connections at deployment time. Multiple MPSs are supported by instantiating multiple SMU with TEI. To support system parameter interpretation across vendors, a metric packet format having three fields is defined: protocol code, unique identifier, and metric value. For efficient hardware fault detection, system parameters values are represented as 32-bit fixed-points, enabling comparators with lower resource utilization and latency compared to floating-point implementations.

The SMU is entirely designed as a hardware architecture to run on an FPGA.
\begin{figure}[!ht]
\centering
\includegraphics[width=0.5\textwidth]{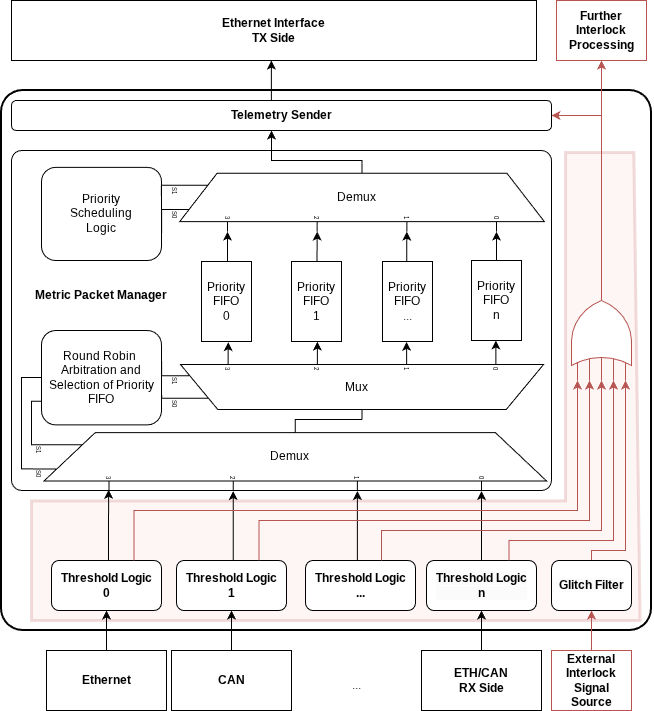}
\caption{SMU microarchitecture, highlighting the interlock logic (red) and the metric packet manager blocks. Black arrows denote AXI-Stream interfaces.}
\label{fig:dataConcentrator}
\end{figure}
As shown in Fig.~\ref{fig:dataConcentrator}, it includes two main components: the interlock logic instances (red block) and the metric packet manager. Here, all modules communicate over AXI-stream interface protocol allowing easy extensions. To  reduce resources for the SMU and avoid high Intellectual Property (IP) costs and its design complexity, the Ethernet stack has been offloaded from FPGA to a dedicated Ethernet module, where data are sent as Metric packets over SPI protocol.

At the microarchitectural level, the SMU evaluates collected system parameters using threshold logic that continuously compares measured values against locally predefined thresholds set at synthesis time. When a system parameter exceeds its corresponding threshold, an error condition is detected and a kill-switch mechanism is activated, asserting the signal high (i.e., logic value 1).
This signal is propagated through OR logic to trigger the following actions: forwarding an alert to users through the Telemetry sender; generating and propagating an external global alert signal to enforce a safe shutdown. All these operations are implemented using combinatorial logic, resulting in a ultra-low latency solution, where the time between error detection in the threshold logic and propagation of the external global alert signal is within a clock cycle, while the latency associated with user notification via the GDV interface depends on network conditions.
In addition to internal monitoring, the SMU accepts external interlock signals from other SMUs. These signals are processed through a glitch filter to suppress spurious transitions, and the resulting global interlock signal is provided as output for subsequent system-level handling.

System parameters values processed by the threshold logic are sent to the Metric Packet Manager over the AXI-Stream interface, even when no errors occur. Since packets from multiple threshold units can arrive simultaneously, a demultiplexer–multiplexer pair, controlled by a round-robin arbiter, selects valid packets, extracts their priority, and routes them to the appropriate priority First-In-First-Out (FIFO) buffer.
The design includes eight priority FIFOs, corresponding to eight severity levels, with priority level 0 representing the highest priority. These FIFOs separate system parameter values, according to the predefined priority rules, provide buffering, and facilitate eventual clock-domain decoupling before packets are forwarded by the Telemetry sender. In addition, arbitration logic within the Metric Packet Manager ensures that metric packets are enqueued into their corresponding priority FIFO and remain buffered until an Ethernet module becomes available to transmit them to the Metric Packet Server, according to the packet priority.
As long as the Telemetry sender is not already sending a UDP alert packet, it is always ready to send Metric Packets while the Ethernet TX Interface (TX-side W5500) has to be ready to send data.

The remaining entities are designed as software modules to run on a Linux operating system.
The MPSrv has three responsibilities: receiving UDP packets sent by the TEI connected to the SMU, converting signed 32-bit fixed-point values into 32-bit floating-point values, and exposing the collected system parameters over HTTP to Prometheus, which scrapes them at regular time intervals. The data stored in the Prometheus database are fetched from the GDE entity.
Although Prometheus does not define a minimum scrape interval, an interval of \SI{10}{\second} was selected for the MRI use case, as each scrape may retrieve thousands of system parameters from the MPSrv. This choice supports continuous system operation while limiting long-term CPU load.
The GDE enables selecting and visualize system parameters using various panel types such as bar graphs, line charts, gauges, and tables, enabling interactive visualization of time-series data.



\section{Evaluation of the GateMate FPGA and its open-source design flow}

This section reports the design experience and platform-specific optimizations required to implement the proposed SMU architecture on the Cologne Chip GateMate M1A1 FPGA using an open-source design flow. The discussion focuses on architectural adaptations, memory utilization strategies, and workflow considerations encountered during migration from an AMD FPGA design to the GateMate platform.

The proposed architecture reused a VHDL controller for the Ethernet transceiver interface, which was originally developed for the AMD 7-series FPGA family based on the W5500 chip. Then, when the project was ported to the GateMate platform, Xilinx’s proprietary AXI4-Stream standard FIFO IP cores were replaced with synthesizable HDL implementations of the FIFOs. System clock generation was also adapted by instantiating a Phase-Locked Loop (PLL) IP core, named CC\_PLL, at the top level of the design.
 
In our design the main bottleneck was the number of BRAM available in the GateMate FPGA. Therefore, analysis of BRAM utilization was done in relation to the datawidth selected, for which we set FIFOs to 20 bits alignments and 1024 depth. While AMD provides AXI-Stream FIFO IP core that can be instantiated in the block design, custom BRAM-based AXI-Stream FIFOs were developed for the GateMate FPGA, according to the vendor BRAM template. In addition, it has been experienced that the GateMate FPGA provides built-in error checking and correction (ECC), which can only be set by instantiating BRAM directly as CC\_BRAM\_20K or CC\_BRAM\_40K primitives with supported data widths (40 or 80 bits) and using 32 or 64 bit words.

For instance, to store 32-bit data words, 40-bit BRAM configuration is used by Yosys, remaining 8 bits to be allocated to Hamming-code parity; In this way, it is possible to enable single-bit error correction and double-bit error detection. ECC parity generation and checking are performed automatically in hardware, introducing a one-clock-cycle data-valid latency and providing dedicated status signals. However, ECC cannot be inferred from generic behavioral VHDL and requires explicit BRAM instantiation with supported data widths.

During migration to the open-source toolchain from AMD FPGA design, behavioral differences were observed in error interpretation and synthesis steps between Vivado and Yosys–GHDL, which may use a different syntax tree interpretation, a relaxed interpretation of the VHDL standard and synthesis strategies. Therefore, HDL previously developed for other platforms should be verified to identify issues related to library and IP core dependencies, as well as race conditions elaborated differently by the two tools. For this reason, in the GateMate workflow, testbenches are executed and waveforms are analyzed with GTKWave, as equivalent behavior cannot be guaranteed across toolchains.

To optimize the design workflow at synthesis, implementation and programming time, starting from the adapted VHDL, GHDL was performed using relaxed semantic checks, and Logic synthesis was executed using the \texttt{synth\_gatemate} flag in Yosys. In addition, we used \texttt{-nomx8} and \texttt{-luttree} flags to disable 8-input multiplexers and to enable LUT tree mapping, respectively. The resulting \texttt{design.json} netlist is inspected to verify correct BRAM inference.
Place and route was performed using nextpnr-himbaechel targeting the \texttt{CCGM1A1} device. Placement may be seeded or randomized, while routing employs a heap-based algorithm to minimize timing slack and meet the PLL-specified system clock frequency. The mandatory \texttt{--router=router2} option was extended with \texttt{--router2-tmg-ripup} to enable timing-driven rerouting, and \texttt{--fpga\_mode=speed} is selected to favor higher achievable clock frequencies over reduced power consumption.
The implementation produces a human-readable output passed to GMpack for bitstream generation, enabling CRC configuration and SPI mode selection. Selecting \texttt{--spimode=quad} exploits the available SPI bandwidth, reducing reconfiguration time by approximately a factor of four compared to single-SPI mode. Bitstreams were uploaded to the GateMate FPGA Starter Kit using openFPGAloader via USB/JTAG or SPI.

Finally, based on the experience gained in this work, a Makefile is provided in the github repository to automate synthesis, implementation, bitstream generation, and programming, improving reproducibility and ease of adoption. Using this workflow, the open-source toolchain achieved shorter synthesis and implementation times than Vivado 2025.2, while Vivado used fewer LUTs, although LUT counts cannot be compared on a one-to-one basis, since the CPE LUT architecture differs from that of the AMD 7-Series family. Recent OSS-CAD-Suite releases further enhance the workflow with CRC-based bitstream validation, Quad-SPI support, and more compact outputs, reducing FPGA reconfiguration time.
\section{Validation \& Evaluation}
To validate the FPGA-centric SCIS architecture and evaluate its performance on the GateMate M1A1 FPGA, a prototype was deployed to be integrated into an ultra-high-field MRI system. The setup includes an MPS that generates synthetic parameters at varying priorities over CAN and a W5500 Ethernet module connecting multiple MPSs to the SMU (hosted on a GateMate Starter Kit). A second W5500 module connects the SMU to the MPSrv, Prometheus, and GDE entities, which run on a Linux workstation (Kernel 6.18).

The hardware design was successfully implemented on the GateMate M1A1 FPGA, which operates at a system clock frequency of \SI{40}{MHz}, constrained by the controller designed into FPGA for the W5500 Ethernet. In practice, data rate of metric packets achieves approximately 18 Mbit/s, which is not a bottleneck for our application. In fact, a single module can process up to 100.000 metric packets per second using the proposed application protocol. A test setup based on a 400 MHz STM32H7ZI2 Nucleo board running FreeRTOS achieved up to 10.000 packets per second with same processing flow. Furthermore, the system is extensible; adding a second W5500 module requires an additional SPI master and UDP packet adapter, consuming only 7\% more LUT resources and three physical BRAM tiles.
As shown in Table \ref{postImplUtilizationComparisonTable}, the SMU architecture utilizes only 37,6\% of LUTs, 9,4\% of FFs, and 35,9\% of BRAM. This low resource footprint facilitates further architectural expansion on the GateMate M1A1 platform. These results were obtained using the workflow and optimization steps previously described within OSS-CAD-Suite v1.13.

\begin{figure}[h]
\centering
\begin{minipage}{0.63\columnwidth}
  \centering
  \includegraphics[width=\linewidth]{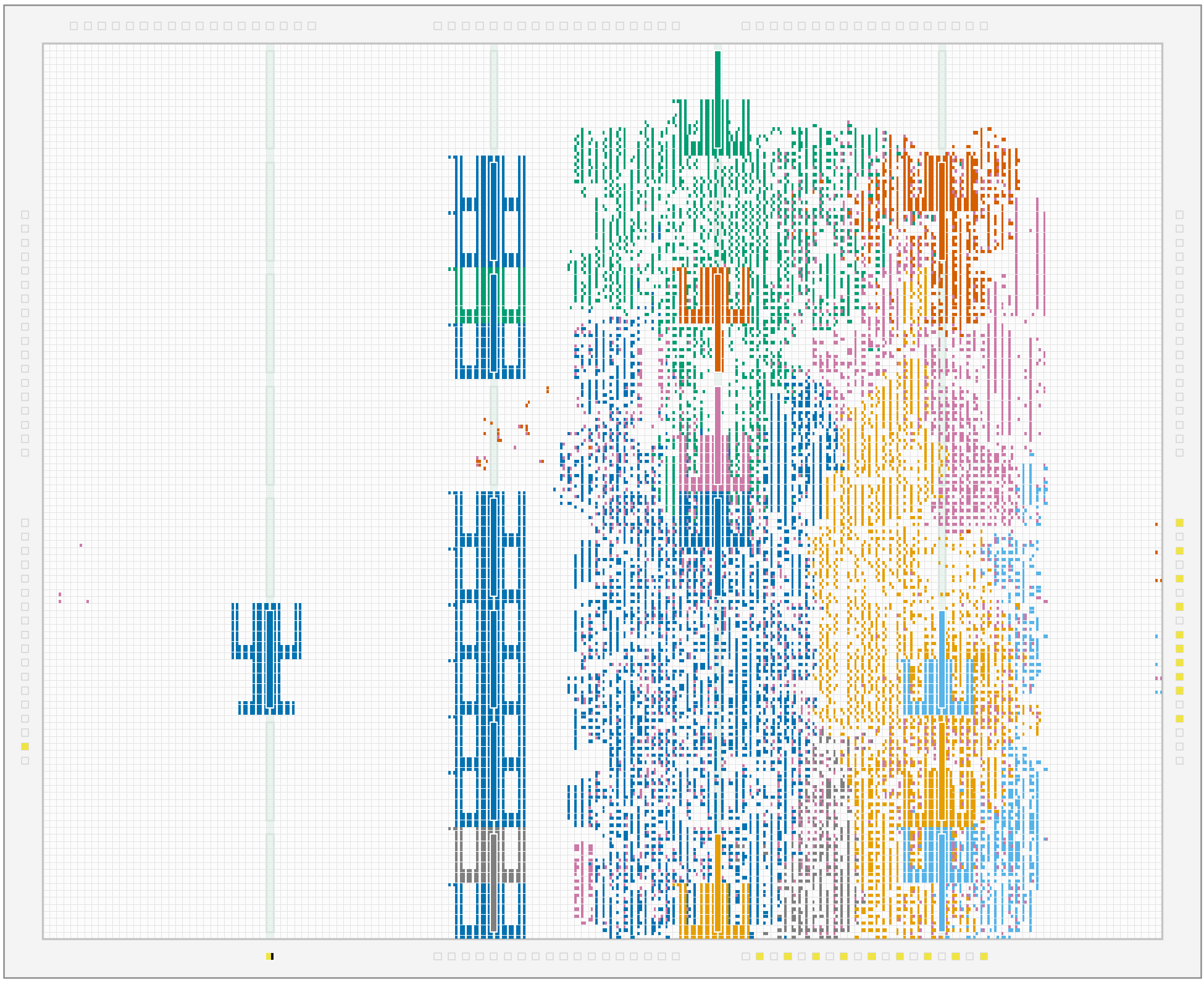}
\end{minipage}
\hfill
\begin{minipage}{0.35\columnwidth}
  \centering
  \includegraphics[width=\linewidth]{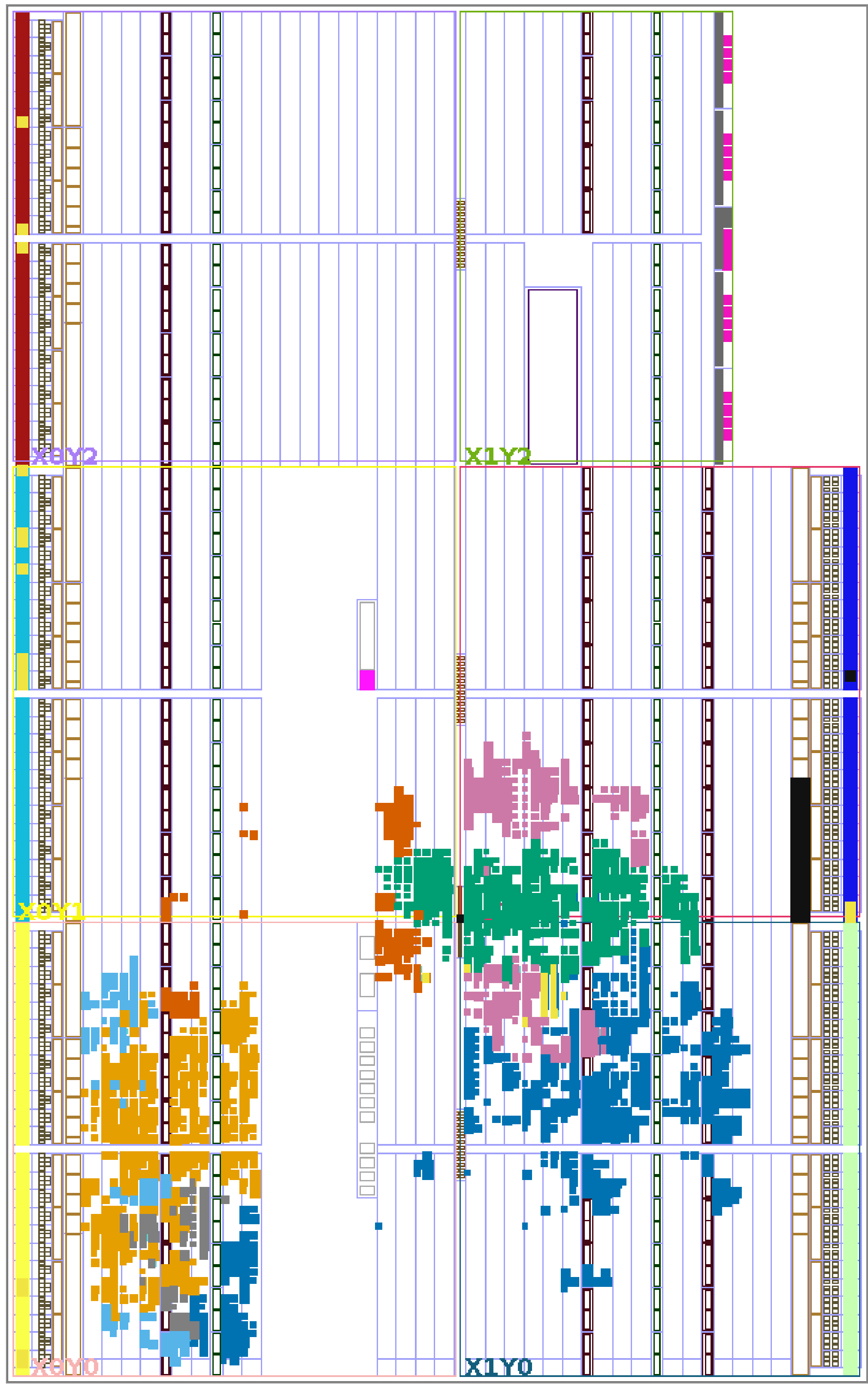}
\end{minipage}
\caption{Post-routing placement GateMate M1A1 (\textit{left}), AMD xc7a35t (\textit{right}).}
\label{fig:placedUtilizationVisual}
\vspace{-2mm}
\end{figure}

\begin{table}[!ht]
\centering
\caption{Post-implementation resource utilization comparison.}
\begin{tabular}{l r r r r r r}
\toprule
\multirow{2}{*}{\textbf{Resource}} 
& \multicolumn{3}{c}{\textbf{GateMate M1A1 FPGA}} 
& \multicolumn{3}{c}{\textbf{AMD Artix-7 xc7a35t}} \\
\cmidrule(lr){2-4} \cmidrule(lr){5-7}
& \textbf{Used} & \textbf{Total} & \textbf{Util. (\%)} 
& \textbf{Used} & \textbf{Total} & \textbf{Util. (\%)} \\
\midrule
GPIO       & 20   & 162    & 12,3  & 20   & 210    & 9,5  \\
LUT        & 15394 & 40960  & 37,6  & 3366 & 20800  & 16,2  \\
FF         & 3833 & 40960  & 9,4   & 3072 & 41600  & 7,4   \\
BRAM       & 23   & 64     & 35,9  & 10,5   & 50     & 21,0  \\
\bottomrule
\end{tabular}
\vspace{3pt}
\begin{minipage}{0.95\linewidth}
\vspace{2pt}
\footnotesize
\textit{Note:} GateMate has 4-input LUTs (LUT4)~\cite{colognechip_gatemate1_datasheet}, whereas
Artix-7 has 6-input LUTs (LUT6). Therefore, LUT counts
are not directly comparable.
\end{minipage}
\vspace{-3mm}
\label{postImplUtilizationComparisonTable}
\end{table}
Table~\ref{postImplUtilizationComparisonTable} compares the resource utilization against the AMD 7-series xc7a35t FPGA. Although, it was previously observed that Vivado is slower than Yosys during synthesis and implementation, from the comparison the SMU architecture results in a lower resource utilization on the AMD device. Furthermore, whereas the design on the xc7a35t was mapped around the three main blocks: the Metric Packet Manager (green), the Threshold Logic, including receivers (cyan), and the Telemetry sender, including Transmitters (red); the Yosys distributed the resources across the GateMate fabric, specifically surrounding the BRAM columns, as shown in Fig.~\ref{fig:placedUtilizationVisual}.

In addition to resource utilization, a key evaluation metric is the reconfiguration time of the architecture running on FPGA. It allows measuring the system-restart time, as rejuvenation method to mitigate aging in the long run. To measure it, the bitstream packed by GMpack 1.13 has been uploaded to the SPI flash memory on the GateMate M1A1 evaluation board and FPGA reconfiguration times were measured with the SIGLENT SDS2104X oscilloscope~\cite{SiglentSDS2000XPlus2021}, starting at the de-assertion of the \texttt{RESET} signal until the rising edge of \texttt{CFG\_DONE} signal.
We observed a reconfiguration time of \SI{40,48}{\milli \second} setting Quad-SPI, achieving an improvement of a factor of 5 compared to an initial implementation, where GMpack 1.2 was used and only SPI single mode was available.
This improvement is not solely due to the SPI mode used for programming the FPGA — the size of the bitstream plays an equally important role. The gmpack v1.13 bitstream generator compresses the BRAM initialisation data, which is written during loading if the memory is not fully utilised. This reduces the number of Half-BRAM CC\_BRAM\_20K cells instantiated in the project. The resulting bitstream is smaller, which further reduces both reconfiguration time and device boot time.

Another key performance metric is the interlock assertion time, defined as the delay to propagate a kill-switch action through the system. Unlike software-based solutions, the architecture is designed for ultra-low-latency response. The interlock is asserted one clock cycle after the last byte is received by the Threshold logic, corresponding to \SI{25}{\nano\second} at \SI{40}{\mega\hertz}. That is in the case of Ethernet communication, which is the slowest interface in the system, the kill-switch is asserted with a delay of about \SI{3}{\micro\second}, while in the best case of an external interlock signal it reaches the aforementioned \SI{25}{\nano\second}. When the SMU receives an interlock assertion as input because it was propagated by another SMU, the interlock assertion is registered and propagated after 4 clock cycles; this latency is introduced by a digital glitch filter with hysteresis behavior reducing unwanted state changes on the global interlock signal inside the SMU.
To propagate the metadata of an asserted kill-switch action to the user over the GDV, the SPI master needs to deserialise data read on the MISO SPI wire to use the W5500 Module. This communication adds 130 clock cycles of latency after the first valid byte of the metric packet data is provided to the Threshold logic.


As part of the system-level validation, the round-trip time (RTT) of Metric Packets was evaluated to characterize the end-to-end communication latency between the FPGA-based SMU architecture and a software-based reference implementation. Given the microsecond-scale timing resolution required, non-deterministic effects introduced by the Linux networking stack were minimized by disabling RX interrupt coalescing at the network interface controller and executing the Python-based dummy SMU on the Pine A64 SBC with real-time scheduling priority and CPU core pinning.  A total of 65.536 RTT samples were collected from the FPGA-based SMU operating at a system clock frequency of \SI{40}{MHz}.

\begin{figure}[!ht]
\centering
\begin{minipage}{0.94\columnwidth}
  \centering
  \includegraphics[width=\linewidth]{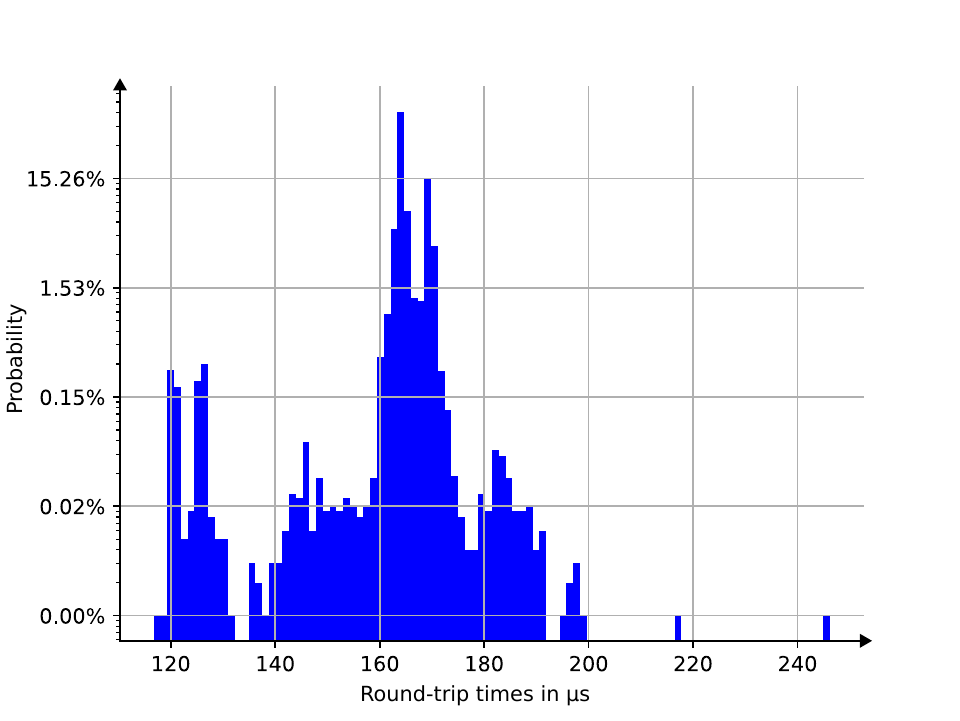}
\end{minipage}
\vspace{-2mm}

\caption{Round-trip time (RTT) measurements.}
\label{fig:FPGAloopbackDelaysHistogram}
\vspace{-5mm}
\end{figure}

The histogram in Fig.~\ref{fig:FPGAloopbackDelaysHistogram} shows a multi-modal latency distribution, with 62,53\% of samples clustered around \SI{164}{\micro\second}, close to the mean RTT of \SI{165}{\micro\second} and a standard deviation of \SI{5}{\micro\second}. The observed jitter, appearing as discrete steps of approximately \SI{10}{\micro\second}, is primarily attributed to Linux interrupt handling and kernel scheduling effects, Python interpreter overhead in the software-based SMU, and the sequential polling of RX socket buffers in the W5500 Ethernet controllers. Despite these sources of variability, the FPGA-based SMU architecture, operating without an operating system layer, achieves consistently low latency, with a maximum measured RTT of \SI{246}{\micro\second}, remaining well below one millisecond and therefore suitable for millisecond-precision time-stamping and timely propagation of UDP-based interlock alert messages.

\section{Conclusions}

This paper presents an FPGA-centric SCIS, designed to provide deterministic system-health monitoring and alert systems, as well as machine protection, using a Grafana dashboard as the operator interface. This is intended for use in long-running safety-critical systems. By moving interlock decision logic and SCS entirely into hardware, the architecture eliminates software stack delays and enables microsecond-scale interlock assertion, while maintaining sub-millisecond end-to-end communication latency. In addition, with offloading Ethernet stack to external W5500 transceiver and toolchain optimizations, the architecture on the GateMate M1A1 FPGA achieves low resource utilization and fast FPGA reconfiguration time of 40,48 ms, enabling architecture extendability, and system rejuvenation as an effective strategy to mitigate aging effects commonly observed in software-based solutions.
Another key contribution of this work is the practical evaluation of the European GateMate M1A1 FPGA and its design flow within the Yosys open-source toolchain. Porting the design required replacing vendor-specific IP with portable VHDL and adapting clocking and memory primitives.
This results in a system that combines the ultra-low latency and deterministic locking behavior of hardware implementations with the scalability of software based SCS, which previously required two separate systems. Results also demonstrates that GateMate can be suitable for a cost-effective, scalable platform for deterministic interlock systems that ensures reliable operation and allows for future architectural extensions in complex applications, such as ultra-high-field MRI.

\section*{Acknowledgment}

This publication was co-funded by the European Union (ERDF) through the project “Universal Integrated Console for Ultra-High-Field Magnetic Resonance Imaging (UIC4UHFMRI)” as part of the program "Sachsen-Anhalt WISSENSCHAFT – Forschung und Innovation (EFRE)".

\bibliographystyle{IEEEtran}
\bibliography{literature}

\end{document}